\documentstyle[preprint,aps]{revtex}
\tighten
\draft
\begin{document}
\title{
Theoretical Physics Institute\\
University of Minnesota\\
\makebox[6in][r]{}\\
\makebox[6in][r]{\tt TPI-MINN-94/28-T}\\
\makebox[6in][r]{\tt August '94}\\
\makebox[6in][r]{}\\
Tunneling into a Two-Dimensional Electron Liquid in a Weak Magnetic
Field}
\author{I. L. Aleiner,$^{(1)}$ H. U. Baranger,$^{(2)}$ and
L. I. Glazman $^{(1)}$}
\address{$^{(1)}$Theoretical Physics Institute, University of
Minnesota, 116 Church Str. SE, Minneapolis, MN 55455\\
$^{(2)}$AT\&T Bell Laboratories,
600 Mountain Ave., Murray Hill, New Jersey 07974-0636}
\maketitle
\begin{abstract}
We study the spectral density function of a two-dimensional
electron liquid in a weak magnetic field, the filling
factor $\nu\gg 1$.  A hydrodynamic model for low-energy excitations
of the liquid is developed. It is found that even at $\nu\gg 1$ the
density of states exhibits a gap at low energies. Its width
$2E_0$ depends on the strength of interaction only logarithmically,
$2E_0=(\hbar\omega_c/\nu)\ln (\nu e^2/\varepsilon\hbar v_F)$.
The effects of temperature and disorder on the density of states
are discussed.
\end{abstract}
\narrowtext
Since the discovery of the integer and fractional quantum Hall
effects\cite{Prange}, the properties of two-dimensional (2D) electron
systems in a strong magnetic field have attracted persistent attention. For a
long time, the magnetotransport coefficients were the main objects of
study. Recently, however, experiments allowing a direct investigation
of the one-electron density of states  were performed
\cite{Ashoori,Eisenstein}. In Ref.~\cite{Eisenstein} the tunnel current
between two 2D layers at small filling factors, $\nu$, was
strongly  suppressed  at low bias, implying the existence of a gap in the
spectral density at low energy.

This fact, which can only be caused by many-electron effects, motivated
extensive theoretical studies\cite{Yang93,Wen,Halperin,Kinaret,Pikus,Wen2} of
tunneling at small filling factors $\nu < 1$. Despite rather different
approaches, all these papers predicted that the width of the gap should
be related to the single length scale in the lowest Landau level by
$E_l\simeq e^2/(\varepsilon l)$ where $\varepsilon$ is the dielectric constant.
The length scale $l$ can be taken to be either the interparticle spacing,
the magnetic length, or the cyclotron radius since these are approximately
equal in this regime.  Such a value of the gap
can be easily understood: $E_l$ equals the Coulomb energy of interaction of
the  extra electron with a ``frozen'' 2D electron system. If there were no
relaxation in the electron system at all, the density of states would vanish
at energies $|\hbar \omega| \lesssim E_l$. The relaxation processes smear out
the threshold in the density of states which therefore becomes sensitive to
the details of these processes.  However, the suppression is still
strong at energies $|\hbar \omega| \lesssim E_l$ if the
characteristic relaxation time exceeds $\hbar/E_l$.

In this Letter we study the energy dependence of the density of states at
large non-integer filling factors, $\nu e^2/(\varepsilon\hbar v_F) \gg 1$
($v_F$ is the Fermi velocity at zero magnetic field). In this regime the
interparticle spacing, the cyclotron radius, and the magnetic length are
distinctly different from each other.  We will show that a
gap followed by a sharp peak exists in the one-electron density of states.
Furthermore, in contrast to the case $\nu \lesssim 1$ the width of the gap
depends only logarithmically on the strength of interaction between electrons.
At $\nu\simeq\varepsilon\hbar v_F/e^2$, which is the lower
boundary of the applicability region, our result approaches the value
$E_l\simeq e^2/\varepsilon l$ with the length scale $l$ equal to the cyclotron
radius.

{\it Qualitative Discussion---}
Our basic approximation is to treat the electrons as a charged
continuous liquid.  Immediately after
the electron tunnels into the electron liquid, the system acquires extra
energy of order $e^2n_e^{1/2}/\varepsilon$ due to the interaction of this
electron with the ones forming the liquid (here $n_e$ is the 2D electron
concentration).
The energy of repulsion is much greater than the energy of the final
state which is determined by an external bias.
Therefore, in order to reach the final
state, the perturbation of the
net electron density has to spread.
The spreading occurs through the virtual emission of
collective excitations, magnetoplasmons, which have a gap equal to
the cyclotron frequency $\omega_c$ in their spectrum:
\begin{equation}
\omega (q)=
\left(\omega_c^2+2\pi |q|\frac{e^2n_e}{\varepsilon m}\right)^{1/2},
\label{plasmon}
\end{equation}
where $m$ is an effective electron mass. Because of the gap, the emission of
magnetoplasmons occurs within a time of order $\omega_c^{-1}$.
Therefore the density perturbation spreads over an area of radius
$a_c \simeq v_g \omega_c^{-1}$ where $v_g$ is the group velocity of the
magnetoplasmon (\ref{plasmon}). Calculating the value of $v_g$ from
Eq.~(\ref{plasmon}), we find
\begin{equation}
 a_c=\frac{2\pi e^2n_e}{\varepsilon m\omega_c^2}.
\label{size}
\end{equation}
As the charge spreads on the time scale $\sim \omega_c^{-1}$,
the Lorentz force causes circular currents to appear.
The Lorentz force due to these
circulating currents
compensates the excess pressure in the region of the density perturbation,
blocking the further spreading of the charge and
making this ``vortex'' configuration stable.

The total energy of the vortex consists of the Coulomb energy plus the kinetic
energy of the circulating currents. The former part is of the order of
$e^2(\varepsilon a_c)^{-1}$. The use of Eq.~(\ref{size}) yields the Coulomb
energy of the vortex $\sim m\omega_c^2/n_e$. Remarkably, this expression does
not contain the gas parameter at all. The kinetic energy can be estimated as
$\sim mv^2(n_ea_c^2)$, where $v$ is the characteristic velocity of the liquid
within the vortex. Using the fact that the Lorentz force compensates the
force of electric repulsion $\sim e^2(\varepsilon a_c^2)^{-1}$, we can write
down the characteristic velocity as $v\sim e^2/(m\omega_ca_c^2)$.
Then, Eq.~(\ref{size}) shows that the kinetic energy approximately
equals the Coulomb energy. The more rigorous treatment below reveals a
logarithmic divergence of the kinetic energy at small distances. This
divergence should be cut off at distances of order the Larmour radius
$R_c=v_F/\omega_c$ since this is the scale of the spatial dispersion
of the kinetic coefficients, and thus the length scale on which
the hydrodynamic approximation
fails. The final result for the energy of the vortex $E_0$ is
\begin{equation}
E_0=\frac{m\omega_c^2}{4\pi n_e}\ln\left(\frac{a_c}{R_c}\right)\equiv
\frac{\hbar\omega_c}{2\nu} \ln\left(\nu
\frac{e^2}{\varepsilon\hbar v_F}\right).
\label{energy}
\end{equation}
This qualitative argument requires the conditions $E_0 \ll
\hbar\omega_c$ and $a_c \gg R_c$ which are both met provided
the inequality $\nu e^2/(\varepsilon\hbar v_F) \gg 1$ holds.

The value given by Eq.~(\ref{energy}) sets a new energy scale associated
with the tunneling electron. If the characteristic time of the further
spreading of the vortex (caused, e.g., by dissipation) is much greater than
$\hbar/E_0$, a sharp peak appears in the spectral density  at the energy
$E_0$.

{\it Electron Liquid Model---}
It is clear from the above discussion that the density of states
at energies $E\lesssim E_0$ is determined by the long wave length
excitations of the electron system. We describe these excitations by a
hydrodynamic model defined by the velocity
$\mbox{\boldmath $v$}(\mbox{\boldmath $r$},t)$ and
electron density $n(\mbox{\boldmath $r$},t)$.
The deviation of the concentration $\delta n(\mbox{\boldmath $r$},t)$ from
its equilibrium value $n_e(\mbox{\boldmath $r$})$
is  related to $\mbox{\boldmath $v$}(\mbox{\boldmath $r$},t)$
by the linearized continuity
equation
$\delta\dot{n}
 +  \mbox{\boldmath $\nabla_{r}$}\left(n_e \mbox{\boldmath $v$}\right) = 0.$
The equilibrium electron density $n_e$ may depend on the coordinate because
of a long range disorder potential produced by remote impurities. However, the
spatial scale of the disorder potential is assumed to be larger than $R_c$,
so that the hydrodynamic description still holds.

It is convenient to take account of the continuity constraint by
introducing the field of  liquid ``displacements''
$\mbox{\boldmath $u$}(\mbox{\boldmath $r$},t)$,
\begin{equation}
    \mbox{\boldmath $v$}\left(\mbox{\boldmath $r$},t\right)=
    \dot{\mbox{\boldmath $u$}}\left(\mbox{\boldmath $r$},t\right),
    \quad
    \delta n\left(\mbox{\boldmath $r$},t\right)= -
    \mbox{\boldmath $\nabla$}
    \left(n_e
    \left(\mbox{\boldmath $r$}\right)
    \mbox{\boldmath $u$}\left(\mbox{\boldmath $r$},t\right)\right).
\label{displacement}
\end{equation}
The Hamiltonian of the liquid in a magnetic field is
\begin{eqnarray}
 &&H =\int d^2r\left\{\frac{1}{2mn_e\left(\mbox{\boldmath $r$}\right)}
\left(\mbox{\boldmath $p$}-
\frac{mn_e\left(\mbox{\boldmath $r$}\right)\omega_c}{2}
\left[\mbox{\boldmath $z$} \times \mbox{\boldmath $u$}\right] \right)^2 +
\right.\nonumber\\
&&\left.\frac{e^2}{2}\int d^2 r_1
\frac{
\mbox{\boldmath $\nabla_r$}
    \left(n_e
    \left(\mbox{\boldmath $r$}\right)
    \mbox{\boldmath $u$}\left(\mbox{\boldmath $r$}\right)\right)
\mbox{\boldmath $\nabla_{r_1}$}
    \left(n_e
    \left(\mbox{\boldmath $r_1$}\right)
    \mbox{\boldmath $u$}\left(\mbox{\boldmath $r_1$}\right)\right)
}
{\left|
\mbox{\boldmath $r$}-\mbox{\boldmath $r_1$}\right|}
\right\},
\label{Hamiltonian}
\end{eqnarray}
where $\mbox{\boldmath $z$}$ is the unit vector perpendicular to the plane
of the liquid. The operator $\mbox{\boldmath $p$}$ is canonically conjugate
to the operator  $\mbox{\boldmath $u$}$,
$\left[p_i(\mbox{\boldmath $r_1$}), u_j(\mbox{\boldmath $r_2$})\right]
=-i\hbar\delta_{ij}\delta\left(\mbox{\boldmath $r_1$}- \mbox{\boldmath
$r_2$}\right).$
We used the cylindrical gauge for the vector potential in the
kinetic energy term in Eq. (\ref{Hamiltonian}); the last term in the
Hamiltonian is the density-density interaction.

{\it Tunneling Formalism---}
For the quantitative description of tunneling we have to evaluate the spectral
density function\cite{Mahan},
\begin{equation}
A(\omega)=
2\mbox{Re}\int_0^{\infty}dt e^{i\omega t}
\langle
\psi\left(t\right)\psi^{\dagger}\left(0\right) +
\psi^{\dagger}\left(0\right)\psi\left(t\right)
\rangle,
\label{density}
\end{equation}
where $\psi^{\dagger}$, $\psi$ are electron creation and annihilation
operators, respectively,
and $\langle\dots\rangle$ denotes averaging over the Gibbs distribution.
We  treat electrons as spinless fermions.

Now we have to establish the relation between the fermionic operators
$\psi^{\dagger}$, $\psi$ and the hydrodynamic variables $\mbox{\boldmath $u$}$
and $\mbox{\boldmath $p$}$. We consider only the local density of states,
for which $\psi^{\dagger}$ and $\psi$ are taken at $\mbox{\boldmath $r$}=0$,
and adopt the
following\cite{Matveev} approximation for the operator creating an electron
at the origin:
\begin{equation}
\psi^{\dagger}(t) \propto \exp [ i \hat{\Pi}(t)/ \hbar ] ,
\label{creation}
\end{equation}
with the ``shift'' operator $\hat{\Pi}$ defined as
\begin{equation}
\hat{\Pi}(t) = \int\frac{d^2\mbox{\boldmath $r$}}{2\pi r^2}
\left(
\frac{\mbox{\boldmath $r\cdot p$}
\left(\mbox{\boldmath $r$},t\right)}{n_e(\mbox{\boldmath $r$})}
+ \frac{m\omega_c}{2} \mbox{\boldmath $r\cdot$}\left[\mbox{\boldmath
$z$}\times\mbox{\boldmath $u$} (\mbox{\boldmath $r$},t)\right]\right).
\label{Pi}
\end{equation}
The operator Eq.~(\ref{creation}) creates an
electron density  perturbation localized near the origin
and does not excite any motion of the electron liquid:
\[
\left[
\psi^{\dagger}(t), \delta n\left(\mbox{\boldmath $r$},t\right)
\right] =-\delta\left(\mbox{\boldmath $r$}\right) \psi^{\dagger}(t),\quad
[\psi^{\dagger}(t), \mbox{\boldmath $v$}\left(\mbox{\boldmath $r$},t\right)]
=0,
\]
where the velocity operator is $\mbox{\boldmath $v$}(\mbox{\boldmath $r$})
\equiv i [ H,\mbox{\boldmath $u$}(\mbox{\boldmath $r$}) ]/\hbar $.
Because we are considering only the local density of states, our calculation
is strictly applicable to the case of tunneling from a point contact to
a 2D plane; however, because the magnetic field restricts the extent of
the electron--- a plane-wave basis is not suitable, for instance--- we believe
that this calculation is relevant to the experimentally interesting case
of plane to plane tunneling. For the case of the local density of states,
we expect the representation Eq.~(\ref{creation}) to be an accurate
description of the long-wavelength collective effects responsible for
the gap structure in the density of states, while the
effects of the single-particle degrees of freedom should not affect this
structure.

The representation Eq. (\ref{creation}) of the creation operator
$\psi^{\dagger}$ in terms of the shift operator $\Pi (t)$ allows us to
relate the spectral density function $A(\omega)$ to the response function
of the electron liquid. Because the Hamiltonian Eq. (\ref{Hamiltonian})
is quadratic, and
operator (\ref{Pi}) is linear in $\mbox{\boldmath $u$ and $p$}$,
the calculation may be performed in a standard way
\cite{Mahan} and results in
\begin{eqnarray}
 A(\omega) \propto \mbox{Re}\int_0^{\infty}dt e^{i\omega t}
\left(e^{J(t)}+e^{J^\star (t)}\right), \label{A} \\
J(t) = i [ D(t)-D(0) ]/\hbar.\nonumber
\end{eqnarray}
Here we introduced the Green function of the electron liquid
$
 D(t) \equiv -i \langle \hat{T}\hat{\Pi}(t)\hat{\Pi}(0)\rangle / \hbar
$
where $\hat{T}$  stands for the time ordering. The most convenient way to
proceed is to relate $D(\omega)$ to the retarded Green function
$
 D^R(t) \equiv -i \theta(t)
\langle[ \hat{\Pi}(t),\hat{\Pi}(0) ]\rangle / \hbar
$
by the identity \cite{Mahan}
\begin{equation}
D(\omega)=\mbox{Re}D^R(\omega) + i \coth\frac{\hbar\omega}{2k_BT}
\mbox{ Im}D^R(\omega).
\label{id}
\end{equation}

Using the equations of motion for the variables $\mbox{\boldmath $u$}$
and $\mbox{\boldmath $p$}$, one can check that
$\dot{\Pi}= \int d^2\mbox{\boldmath $r$}
\delta n(\mbox{\boldmath $r$})(e^2/\varepsilon|\mbox{\boldmath $r$}|)$.
This would allow one to express the Green functions $D$ and $D^R$
in terms of corresponding density-density correlation functions\cite{Mahan}.
However, it is more convenient to obtain an equation that
determines directly the Green function $D^R$. To derive this equation,
we notice that the retarded Green function coincides with the linear
response of the average
$\langle\hat{\Pi}(t)\rangle$ to an external perturbation of the form
$\delta(t)\hat{\Pi}(t)$ added to the Hamiltonian Eq. (\ref{Hamiltonian}). We
calculate this linear response  with the help of the equations of motion for
$\mbox{\boldmath $u$ and $p$}$.
Because the Hamiltonian Eq. (\ref{Hamiltonian}) is quadratic the equation for
the Green function can be found in a closed form.
A simple calculation yields
$D^R (t)=\lim_{r \to 0}\tilde{D}^R(t,\mbox{\boldmath $r$})$,
where $\tilde{D}^R(t,\mbox{\boldmath $r$})$ is the solution of the equation
\begin{eqnarray}
&&\displaystyle{\frac{e^2}{\varepsilon m}
\left\{
\left(\mbox{\boldmath $\nabla$}
\frac{\partial}{\partial t}+
\omega_c\left[z  \times
\mbox{\boldmath $\nabla$}\right]
\right)
\int {d^2r_1}
\frac{n_e(\mbox{\boldmath $r$}_1)
\mbox{\boldmath $\nabla$}_{r_1}
\tilde{D}^R(\mbox{\boldmath $r$}_1)}
{\left|\mbox{\boldmath $r$}-\mbox{\boldmath $r$}_1\right|}\right\}
}
\nonumber\\
&&\displaystyle{+
\frac{\partial}{\partial t}
\left(
\omega_c^2+\frac{\partial^2}{\partial t^2}\right)
\tilde{D}^R(\mbox{\boldmath $r$})=
-[\omega_c^2\theta(t)+\delta^{\prime}(t)]\frac{e^2}
{\varepsilon|\mbox{\boldmath $r$}|},
}
\label{B}
\end{eqnarray}
at $t \geq 0$, and  $\tilde{D}^R(t,\mbox{\boldmath $r$})=0$ at $t < 0$.

{\it Homogeneous Liquid---}
For a homogeneous liquid ($n_e={\rm const}$), Eq. (\ref{B}) can be easily
solved by Fourier transformation. Because $E_0 \ll \hbar \omega_c$,
only the asymptotic behavior at $t\gg\omega_c^{-1}$
is needed.
Using this solution in Eqs.~(\ref{A}) and (\ref{id}), we
obtain
\begin{equation}
J(t)=\left(-i\frac{E_tt}{\hbar}-\frac{k_BTE_tt^2}{\hbar^2}\right)
\int_0^{1/R_c}dk\frac{a_c}{1+a_ck},
\label{J}
\end{equation}
where $a_c$ is defined by Eq.~(\ref{size}), and
\begin{equation}
E_t=\frac{m\omega_c^2}{4\pi n_e}.
\label{Et}
\end{equation}
The logarithmic divergence of the integral in Eq. (\ref{J}) is cut off at
wavevectors of order $R_c^{-1}$ where the hydrodynamic approximation fails.
Substituting Eq.~(\ref{J}) into Eq.~(\ref{A}) yields the spectral
density for $k_BT\ll E_0$,
\begin{equation}
 A(\omega) \propto \frac{\hbar}{\sqrt{k_BTE_0}}
\exp\left(-\frac{\left(|\hbar\omega| - E_0\right)^2}{4E_0k_BT}\right),
\label{nuhom}
\end{equation}
with energy $E_0$ defined by Eq.~(\ref{energy}). The lower the temperature, the
sharper the peak in the density of states at $\hbar\omega=E_0$. The gap at
smaller energies is due to the blockade of spreading of the electron liquid.
As we already discussed, spreading results in the formation of a finite-size
vortex with energy $E_0$. The peak at negative energy corresponds to an
``antivortex'' formed after an electron tunnels out from the electron liquid.

Finite temperature broadens the peak in the density of states by introducing
some initial inhomogeneities in the electron liquid. For example, a vortex
carrying some charge $\delta q$ can be formed spontaneously by a thermal
fluctuation; the probablity of such a configuration is proportional to
$\exp(-\epsilon_i/k_BT)$ with energy $\epsilon_i=E_0(\delta q/e)^2$.
Tunneling of an electron into the center of this configuration produces a
vortex containing charge $\delta q + e$, and therefore having energy
$\hbar\omega = E_0(1+2\delta q/e)$. These qualitative considerations lead
directly to Eq.~(\ref{nuhom}).

As we already mentioned, the sharp peak in the density of states at
non-zero energy occurs because the charge spreading is blocked by the
magnetic field applied to an ideal, homogeneous electron liquid. There are
several ways in which this result should be modified in the non-ideal case.
First, in a ``viscous'' liquid, a finite longitudinal dc-conductivity
$\sigma_{xx}$ allows charge spreading, and the gap is washed out. The
microscopic theory\cite{Halperin2} of conductivity for a special case
$\nu=1/2$ allowed He {\em et al.}\cite{Halperin} to obtain the function
$A(\omega)$ in a wide range of energies.
Second, at small filling factors the
electron liquid may crystalize into a Wigner crystal. It was shown by
Johansson and Kinaret\cite{Kinaret} that the peak in the tunneling
density of states in this case has a finite width.
This occurs solely due to the existence of gapless magnetophonon
modes in a crystal that are absent in a liquid.
Neither of these approaches is suitable for the
large filling factors considered here.
Third, the charge may spread
because of inhomogeneities \cite{footnote} caused, e. g.,
by  smooth disorder; we pursue this
possibility below.

{\it Inhomogeneous Liquid---}
It is well known that density inhomogeneities may give rise to gapless
excitations of an electron liquid in a magnetic field. These  gapless modes
redistribute charge, and thus cause spreading of the vortex and broadening
of the peak in the density of states. The origin of such excitations is
analogous to that of the edge magnetoplasmon\cite{magnetoplasmon}
propagating along the periphery of the system.

In order to estimate the possible effect of smooth disorder, we start
with the simplest model of a constant gradient of the concentration
$n_e(\mbox{\boldmath $r$})$ in
the region of the liquid into which the electron tunnels:
\begin{equation}
 n_e(\mbox{\boldmath $r$})=\bar{n}_e +
|\mbox{\boldmath $\nabla$} n_e| x.
\label{concentration}
\end{equation}
Here the $x$ axis is in the direction of the gradient,
and we assume the potential to be smooth so that the concentration changes
slowly on the scale set by the vortex size,
${|\mbox{\boldmath $\nabla$} n_e|}a_c \ll \bar{n}_e$.
In this case Eq.~(\ref{B})  for zero temperature and  $\omega_ct \gg 1$ yields
\begin{equation}
J(t)=2\pi E_{t}
\int\frac{d^2k}{(2\pi)^2}\frac{a_c\theta(\omega(\mbox{\boldmath $k$})t)}
{|\mbox{\boldmath $k$}|\left(1+a_c|\mbox{\boldmath $k$}|\right)}
\frac{e^{-i\omega(k)t}-1}
{|\hbar\omega(\mbox{\boldmath $k$})|},
\label{Jnh}
\end{equation}
where $E_t$ is defined by Eq.~(\ref{J}), and the spectrum
$\omega(\mbox{\boldmath $k$})$ of the
magnetoplasmon excitations which are associated with the inhomogeneity
Eq. (\ref{concentration}) is given by
\begin{equation}
\omega(\mbox{\boldmath $k$})= \tau_d^{-1}
\frac{k_y}
{|\mbox{\boldmath $k$}| (1+a_c|\mbox{\boldmath $k$}|)},\quad
\tau_d\equiv\frac{\bar{n}_e}{\omega_c a_c |\mbox{\boldmath $\nabla$} n_e|}.
\label{omega}
\end{equation}

At $t \ll \tau_d$ the exponent in Eq.~(\ref{Jnh}) can be expanded, giving
$J(t)={-iE_0t}/{\hbar} - {E_tt^2}/({2\pi\tau_d\hbar})$, similar to
Eq.~(\ref{J}). The corresponding result for the spectral density function at
energies
$||\hbar\omega|-E_0|\lesssim E_t$ is:
\begin{equation}
 A(\omega) \propto \sqrt{\frac{\hbar\tau_d}{E_t}}
\exp\left[
-\frac{\pi\tau_dE_t}{2\hbar}
\left(
\frac{|\hbar\omega| - E_0}{E_t}
\right)^2\right].
\label{nunonhom}
\end{equation}
Thus, the spreading of the charge through the low-frequency
modes, Eq. (\ref{omega}), leads to a broadening of the peak similar in form to
that caused by temperature in the homogeneous case, Eq. (\ref{nuhom}).
As in the case of finite temperature, one can view this broadening
as due to fluctuations in the initial conditions; however, the fluctuations
now have a quantum origin in that they are produced by the zero-point
motion of the low-frequency modes.

The region of validity of Eq. (\ref{nunonhom}) corresponds to relatively
short times $t\lesssim \tau_d$ contributing to the integral in Eq. (\ref{A}).
On this time scale
the vortex spreads on a distance $l_d\simeq \tau_d (d\omega/dk_y)|_{|k|\simeq
1/a_c}\simeq a_c$. According to our assumption about the smoothness of the
disorder, the corresponding spatial variation of electron density is small,
which justifies the use of the gradient expansion Eq. (\ref{concentration}).

We have found that the width of the peak in the local density of states
is determined by the local concentration gradient in the case of smooth
disorder. Turning now to large-area tunnel barriers,
the differential conductance is related to the ensemble average of the
density of states, and therefore the width of the peak of the average
density of states is determined by the typical value of the
concentration gradient. This yields a relation
between the observable width of the peak and parameters of a sample.

For estimates we assume that the 2D electron liquid is created in a
heterostructure, where the random potential is produced by a layer of the
remote
charged impurities with concentration $N_D$, and that all the electrons in the
2D liquid originate from these impurities, $\langle{n}_e\rangle=N_D$.
Then, the characteristic spatial scale of the disorder potential is equal to
the
spacer width $d$ and the typical deviation of the concentration from its
average
value  $\langle{n}_e\rangle$ is $\sqrt{N_D}/d$. The density fluctuations
are small if $N_Dd^2\gg 1$.
We can determine the width of the peak $\delta E \sim (E_t\hbar/\tau_d)^{1/2}$
after estimating the characteristic values of $E_t$ and $\tau_d$:
\begin{equation}
\frac{\delta E}{E_0}\sim\frac{a_B}{d}
\left(\nu \frac{e^2}{\varepsilon\hbar v_F}\right)^{3/2}
\left[\ln\left(\nu\frac{e^2}{\varepsilon\hbar v_F}\right)\right]^{-1}
\label{width}
\end{equation}
where $a_B=\varepsilon\hbar^2/e^2m$.
As discussed above, this width is caused by quantum fluctuations
of the low-frequency modes of the inhomogeneous electron liquid.
Note that this width is significantly larger than that caused by
the direct effect of static disorder on $E_0$ [i.e., by $n_e (r)$ variation in
Eq. (\ref{energy})].
 The constraint of smooth disorder on the scale
set by the vortex size is satisfied if $\nu\lesssim (\varepsilon\hbar
v_F/e^2)(d/a_B)^{1/2}$. The opposite case of a ``short-range'' potential
$R_c\ll d\ll a_c$ will be considered elsewhere.

In conclusion, we have demonstrated that even relatively weak magnetic field
induces a gap in the tunneling density of states of an electron liquid. For
a homogeneous liquid this gap is followed by a sharp peak at a finite energy.
Low-frequency modes induced by disorder
broaden this peak through
quantum fluctuations.
Eq. (\ref{width}) shows that for a
high-quality
heterostructure ($d/a_B\simeq10$) the peak should be observable for filling
factors $\nu \sim 3 - 4$.

Discussions with  R. C. Ashoori,  J. P. Eisenstein,   B. I. Halperin, K. A.
Matveev  and B. I. Shklovskii are gratefully acknowledged.
The work at the University of
Minnesota was  supported by NSF Grant DMR-9117341.


\begin{references}
\bibitem{footnote} The effect of disorder for $\nu \leq 1/2$ has been
studied numerically:
a model of a classical electron liquid was used in Ref.\cite{Pikus}, and
the Hartree-Fock approximation for the case of a strong disorder was applied
in Ref.\cite{Yang93}
\end{references}
\end{document}